\documentclass[12pt]{iopart}
\usepackage{iopams}
\usepackage{rotating}
\usepackage{epsfig}
\bibstyle{plain}
\newcommand{\be}{\begin{equation}}
\newcommand{\ee}{\end{equation}}
\newcommand{\bea}{\begin{eqnarray}}
\newcommand{\eea}{\end{eqnarray}}

\eqnobysec

\begin{document}
\paper[Thermal Casimir effect]{Thermal corrections to the Casimir effect}
\author{Iver Brevik$^1$, Simen A Ellingsen$^1$ and
Kimball A Milton$^2$\footnote{Permanent address:
Oklahoma Center for High Energy Physics and Homer L. Dodge
Department of Physics and Astronomy, The University of
Oklahoma, Norman, OK 73019 USA}}
\address{$^1$Department of Energy and Process
Engineering, Norwegian University of Science and
Technology, N-7491, Trondheim, Norway\\
$^2$Department of Physics, Washington University, St.~Louis,
MO 63130 USA}\ead{iver.h.brevik@ntnu.no,simenand@stud.ntnu.no,
milton@nhn.ou.edu}

\date\today

\begin{abstract}
The Casimir effect, reflecting quantum vacuum fluctuations in the 
electromagnetic field in a region with material boundaries, has been studied
both theoretically and experimentally since 1948.  The forces between 
dielectric and metallic surfaces both plane and curved have been measured at
the 10 to 1 percent level in a variety of room-temperature experiments, 
and remarkable agreement with the zero-temperature theory has been achieved. 
In fitting the data various corrections due to surface roughness, patch
potentials, curvature, and temperature have been incorporated.  It is the 
latter that is the subject of the present article.  We point out that, in 
fact, no temperature dependence has yet been detected, and that the 
experimental situation is still too fluid to permit conclusions about 
thermal corrections to the Casimir effect.  Theoretically, there are subtle 
issues concerning thermodynamics and electrodynamics which have resulted in
disparate predictions concerning the nature of these corrections.
However, a general consensus has seemed to emerge that suggests that the
temperature correction to the Casimir effect is relatively large, and should
be observable in future experiments involving surfaces separated at the
few micrometer scale.
\end{abstract}

\maketitle

\section{Introduction}
\label{sec:intro}
About the same time that Schwinger and Feynman were inventing renormalized
quantum electrodynamics, Casimir discovered that quantum electrodynamic
fluctuations resulted in macroscopic forces between conductors and 
dielectrics \cite{casimir}. 
The theory was a natural outgrowth of the Casimir-Polder
theory of the retarded dispersion force between molecules \cite{cp}.
The general theory for the forces between parallel dielectrics was worked
out by Lifshitz and collaborators \cite{lifshitz}, who also included
temperature corrections, which were considered further by Sauer \cite{sauer}
and Mehra \cite{mehra}.  Some years later, the whole theory was rederived
by Schwinger, DeRaad, and Milton \cite{schwinger78}.

The early experiments on Casimir forces were rather inconclusive -- for a 
review see \cite{sparnaay}.  However, the corresponding Lifshitz theory was
verified rather impressively by Sabisky and Anderson \cite{sabisky}, so there
could hardly be any doubt of the validity of the essential ideas.  Starting
about a decade ago, modern experiments by Lamoreaux \cite{lamoreaux,
lamoreaux1, lamoreaux2, lamoreaux3}, Mohideen and collaborators
 \cite{mohideen,mohideen1,mohideen2}, and by Erdeth \cite{ederth}
brought the experimental measurement of the Casimir force between
curved metal surfaces (mapped to the plane geometry by the proximity
approximation \cite{blocki,deriagin}) into the percent accuracy region.  
(Exact results have now apparently rendered the use of the proximity
approximation, which cannot be extended beyond leading order, unnecessary.
See, for example, \cite{emig,gies,bordag}.)  Application of such Casimir
forces to nanoelectromechanical devices have been suggested by experiments
at Bell Labs and Harvard \cite{chan, chan1,iannuzzi}. Only one experiment 
so far, of limited accuracy ($\sim15$\%), has employed parallel plates
\cite{bressi}.  The difficulty of maintaining parallelism in that geometry
limits the accuracy of the experiment, but the forces are much larger than
those between a sphere and a plate, so the forces can, in principle, be 
determined at much larger separations.  Proposals to perform measurements
of the force
between a cylinder and a plane \cite{brown-hayes} and between eccentric
cylinders \cite{dalvit} have advantages because the forces are stronger
than between a sphere and a plane, yet the difficulties in assuring
parallelism are not so severe as with two plane surfaces. 
The most precise experiments so
far, based on both static and dynamical procedures between a plate and a
spherical surface, have been performed at Purdue \cite{decca,decca1,decca2},
where the accuracy is claimed to be better than 1\% at separations down to
less than 100 nm.  

All present experiments agree well with the zero-temperature
Casimir theory when surface roughness and finite conductivity 
corrections are included \cite{bordagrev,lamoreauxrev}.  
The issue about which controversy
has recently erupted is the temperature dependence.  (For recent statements
of both sides of the controversy, see \cite{decca2, Brevik:2004ue, 
Bezerra:2005hc,  Hoye:2005eu, most05}.)  All experiments reported to date  have
been conducted at room temperature, so there is no direct evidence for or 
against any particular model of the temperature dependence.  Indirect evidence
for this dependence has been inferred based on the nonzero shift in the 
theoretical Casimir force between the surfaces due to the difference between
the force at zero temperature and at 300 K.  Surprisingly, this temperature
shift is not so straightforwardly computed as one would have at first 
suspected.

It is the purpose of the present paper to frame the question of the
temperature dependence of the Casimir force in the
context of the history of the subject and the present experimental constraints,
as well as to point out ways
of reconciling the ambiguities both from the theoretical and experimental
sides.  In the following section we review the standard approach given in
\cite{schwinger78} for both dielectric and metal surfaces.
Then, in \sref{Sec:te0} we give the arguments why the transverse electric
(TE) zero mode should not be included, and how this impacts the temperature
dependence of the force, and the resulting impact on the free energy and
entropy.  Other theoretical arguments for and against this point of view
are discussed in \sref{Sec:arg}.  The status of the experimental
situation, and the possibility of dedicated experiments to search for the
temperature dependence of the Casimir effect, will be reviewed in 
\sref{Sec:exp}.  Some new calculations are presented in 
\sref{sec:calc} in the hope of providing signatures to help resolve the 
controversy. Finally, concluding remarks are offered in \sref{Sec:Concl}.

\section{Conventional temperature approach}
\label{Sec:conv}

The zero-temperature Casimir effect between parallel conducting plates,
or between parallel dielectrics, is very well understood, and is not 
controversial.  The formula for the latter, which includes the former as a
singular limit, may be derived by a multitude of formalisms, which will not
be reviewed here \cite{schwinger78,bordagrev,miltonbook,
Hoye:2002at,Milton:2004ya}.
For a system of parallel dielectric media, characterized by a permittivity
\be
\varepsilon(z)=\left\{\begin{array}{cc}
\varepsilon_1,&z<0,\\
\varepsilon_3,&0<z<a,\\
\varepsilon_2,&a<z,
\end{array}\right.
\ee
where the various permittivities are functions of frequency, the 
Lifshitz force per unit area on one of the surfaces is at zero temperature
\be
P^{T=0}=-\frac1{4\pi^2}\int_0^\infty \rmd \zeta\int_0^\infty \rmd k_\perp^2
\kappa_3(d^{-1}+d^{\prime -1}),\label{lifshitz0t}
\ee
where $\zeta$ is the imaginary frequency, $\zeta=-\rmi\omega$, and the
longitudinal wavenumber is 
\be
\kappa_i=\sqrt{k_\perp^2+\zeta^2\varepsilon_i(\rmi \zeta)},
\ee
while the transverse electric (TE) and transverse magnetic (TM) Green's 
functions are characterized by the denominators
\be
d=\frac{\kappa_3+\kappa_1}{\kappa_3-\kappa_1}\frac{\kappa_3+\kappa_2}{\kappa_3
-\kappa_2}\rme^{2\kappa_3 a}-1,\qquad
d'=\frac{\kappa'_3+\kappa'_1}{\kappa'_3-\kappa'_1}\frac{\kappa'_3+\kappa'_2}
{\kappa'_3-\kappa'_2}\rme^{2\kappa_3 a}-1,\label{denominators}
\ee
respectively, where $\kappa_i'=\kappa_i/\varepsilon_i$.

The attractive Casimir pressure between parallel perfectly conducting planes 
separated by a vacuum space of thickness
$a$ is obtained by setting $\varepsilon_{1,2}\to\infty$ and $\epsilon_3=1$.
In that case the TE and TM contributions are equal, and we have
\be
\fl P_C=-\frac1{8\pi^2}\int_0^\infty \rmd \zeta
\int_{\zeta^2}^\infty \rmd \kappa^2\frac{4\kappa}{\rme^{2\kappa a}-1}
=-\frac1{\pi^2}\int_0^\infty \rmd \zeta\frac{\zeta^3}{\rme^{2\zeta a}-1}
=-\frac{\pi^2\hbar c}{240 a^4},\label{caszero}
\ee
which is Casimir's celebrated result \cite{casimir}.

The controversy surrounds the question of how to incorporate thermal 
corrections into the latter result.  At first glance, the procedure to do
this seems straightforward.  It is well-known that thermal Green's functions
must be periodic in imaginary time, with period $\beta=1/T$ 
\cite{Martin:1959jp}.  This implies a Fourier series decomposition, rather 
than
a Fourier transform, where in place of the imaginary frequency integral
in \eref{caszero} we have a sum over Matsubara frequencies
\be 
\zeta_m^2=\frac{4\pi^2 m^2}{\beta^2},
\ee
that is, the replacement
\be
\int_0^\infty \frac{\rmd\zeta}{2\pi}\to\frac1\beta\sum_{m=0}^\infty{}',
\label{finitetsum}
\ee
the prime being an instruction to count the $m=0$ term in the sum with half
weight.  This prescription leads to the following formula for the Casimir
pressure between perfect conductors at temperature $T$,
\be
P^T=-\frac1{4\pi\beta a^3}\sum_{m=0}^\infty{}'\int_{mt}^\infty y^2\,\rmd y
\frac1{\rme^y-1},\label{perfectt}
\ee
where
\be
t=\frac{4\pi a}\beta.
\ee
From this it is straightforward to find the high and low temperature
limits,
\numparts
\bea
P^T\sim -\frac1{4\pi\beta a^3}\zeta(3)-\frac1{2\pi\beta a^3}\left(1+t+
\frac{t^2}2\right)\rme^{-t},\qquad \beta\ll 4\pi a,\label{hight}\\
P^T\sim-\frac{\pi^2}{240 a^4}\left[1+\frac{16}3\frac{a^4}{\beta^4}
-\frac{240}{\pi}\frac{a}\beta\rme^{-\pi\beta/a}\right],\qquad 
\beta\gg 4\pi a.\label{lowt}
\eea
\endnumparts
These are the results found by Sauer \cite{sauer} and Mehra \cite{mehra},
and by Lifshitz \cite{lifshitz}.  The two limits are connected by the
duality symmetry found by Brown and Maclay \cite{brown}.
The pressure may be obtained by differentiating the free energy,
\be
P=-\frac\partial{\partial a}F,
\ee
which takes the following form for low temperature (now omiting the 
exponentially small terms)
\be
F\sim -\frac{\pi^2}{720 a^3}-\frac{\zeta(3)}{2\pi}T^3+\frac{\pi^2}{45}
T^4a,\qquad aT\ll1,
\ee
from which the entropy follows,
\be
S\sim-\frac\partial{\partial T}F\sim \frac{3\zeta(3)}{2\pi}T^2-
\frac{4\pi^2}{45}T^3a,\qquad aT\ll1,
\ee
which vanishes as $T$ goes to zero, in accordance with the third law of
thermodynamics, the Nernst heat theorem.

\section{Exclusion of TE zero mode}
\label{Sec:te0}

However, there is something peculiar about the procedure adopted above for a
perfect metal. (This seems first to have been appreciated by
Bostr\"om and Sernelius \cite{bostrom}.)
 It has to do with the transverse electric mode of zero 
frequency,
which we shall refer to as the TE zero mode.  If we examine the zero frequency
behavior of the reflection coefficients for a dielectric appearing in 
\eref{denominators},
we see that providing $\zeta^2\varepsilon(\rmi \zeta)\to0$ as $\zeta\to0$, the
longitudinal wavenumber $\kappa_i\to k$ as $\zeta\to0$, and hence $d\to\infty$
as $\zeta\to0$.  This means that there is no TE zero mode for a dielectric.
This statement does not appear to be controversial \cite{geyer05}.  
However, if a metal is modeled as the $\varepsilon\to\infty$ limit of a 
dielectric, the same conclusion
would apply.  Because that would spoil the concordance with the third law noted
in the previous section, the prescription was promulgated in \cite{schwinger78}
that the $\varepsilon\to\infty$ limit be taken before the $\zeta\to0$ limit.
But, of course, a real metal is not described by such a mathematical limit, so 
we must examine the physics carefully.

A simple model for the dielectric function is the plasma dispersion relation,
\be
\varepsilon(\omega)=1-\frac{\omega_p^2}{\omega^2},\label{plasma}
\ee
where $\omega_p$ is the plasma frequency.  For this dispersion relation,
the condition $\zeta^2\varepsilon(\rmi\zeta)\to0$ fails to hold as $\zeta\to0$,
and the idealized prescription result, namely the contribution of the TE
zero mode, holds.  However, real metals are not well described by this
dispersion relation.  Rather, the Drude model,
\be
\varepsilon(\rmi \zeta)=1+\frac{\omega_p^2}{\zeta(\zeta+\gamma)},
\label{drude}
\ee
where the relaxation frequency $\gamma$ represents dissipation, very
accurately fits optical experimental data for the permittivity for
$\zeta  < 2\times 10^{15} $ rad/s \cite{lambrecht, lambrecht2}.
For example, for gold, appropriate values of the parameters are \cite{ba}
\be
\omega_p=9.03 \mbox{ eV},\qquad \gamma=0.0345 \mbox{ eV}.
\ee
In this case, the arguments given above for the exclusion of the TE
zero mode apply.

The arguments are a bit subtle \cite{Hoye:2002at, Milton:2004ya}, so we
review and extend them here.  Let us write the Lifshitz formula 
at finite temperature in the form
\be
P^T=\sum_{m=0}^\infty{}'f_m=\int_0^\infty \rmd m\,f(m)-\sum_{k=0}^\infty
\frac{B_{2k}}{(2k)!}f^{(2k-1)}(0),
\ee
where the second equality uses the Euler-Maclaurin sum formula, in terms
of
\be
f(m)=-\frac1{2\pi\beta}\int_0^\infty \rmd k_\perp^2\,
\kappa(\zeta_m)\left(d_m^{-1}+d_m^{\prime-1}\right),
\label{fm}
\ee
according to \eref{lifshitz0t} and \eref{finitetsum}, where we assume
that vacuum separates the two plates so $\kappa_3(\zeta_m)=
\kappa(\zeta_m)=\sqrt{k_\perp^2+\zeta_m^2}.$  Here the denominators 
\eref{denominators}
are functions of $\zeta_m$.  By changing the integration variable from
$m$ to $\zeta_m$ we immediately see that the integral term in the
Euler-Maclaurin sum formula corresponds precisely to the zero-temperature
result \eref{lifshitz0t}.  

One must, however, be careful in computing the low temperature corrections
to this.  One cannot directly expand the denominator $d$ in powers of $\zeta$
because the $k_\perp$ integral in \eref{fm} ranges down to zero.  Let us 
rewrite the TE term there as follows:
\be
\fl 
f^{({\rm TE})}(m)=-\frac1{\pi\beta}\int_{2m\pi/\beta}^\infty \rmd \kappa\, 
\kappa^2
\left\{\left[\frac{1+\sqrt{1+\zeta_m^2(\varepsilon(\rmi\zeta_m)-1)/\kappa^2}}
{1-\sqrt{1+\zeta_m^2(\varepsilon(\rmi\zeta_m)-1)/\kappa^2}}\right]^2
\rme^{2\kappa a}-1\right\}^{-1}.\label{tesummand}
\ee
Evidently, for the Drude model, or more generally, whenever
\be
\lim_{\zeta\to0}\zeta^2[\varepsilon(\rmi \zeta)-1]=0,
\ee
$f^{({\rm TE})}(0)=0$.  However, it is important to appreciate the physical
discontinuity between $m=0$ and $m=1$ for room temperature.  At 300 K,
while $\zeta_0=0$, $\zeta_1=2\pi T=0.16$ eV, large compared the 
relaxation frequency $\gamma$.  Therefore, for $m>0$,
\bea
f^{({\rm TE})}(m)\approx-\frac1{\pi\beta}\int_{\zeta_m}^\infty \rmd \kappa\,
\kappa^2\left[\left(\frac{\sqrt{1+\omega_p^2/\kappa^2}+1}
{\sqrt{1+\omega_p^2/\kappa^2}-1}\right)^2\rme^{2\kappa a}-1\right]^{-1}
\nonumber\\
\approx-\frac1{\pi\beta}\int_{\zeta_m}^\infty \rmd \kappa\,\kappa^2
\frac1{\rme^{2\kappa a}-1},
\eea
provided the significant values of $\zeta_m$ and $\kappa$ are small
compared the the plasma frequency $\omega_p$.  This is just the ideal
metal result contained in \eref{perfectt}.  Insofar as this is accurate,
this expression yields the low- and high-temperature corrections seen in
\eref{lowt}, \eref{hight}.  However, there is now a discontinuity in
the function $f^{({\rm TE})}$. As $\zeta_m\to0$,
\be
f^{({\rm TE})}(m)\to-\frac1{\pi\beta}\int_0^\infty \rmd \kappa\frac{\kappa^2}{
\rme^{2\kappa a}-1}=-\frac{\zeta(3)}{4\pi\beta a^3},
\ee
rather than zero.  This implies an additional linear term in the pressure
at low temperatures:
\be
P^T\sim P^{T=0}+\frac{\zeta(3)}{8\pi a^3}T, \qquad aT\ll1.\label{linear}
\ee
Exclusion of the TE zero mode will also reduce the linear temperature
dependence expected at high temperatures,
\be
P^T\sim -\frac{\zeta(3)}{8\pi a^3}T,\qquad aT\gg1,\label{hightlinear}
\ee
one-half the usual ideal metal result seen in \eref{hight}, 
and this is indeed predicted in
numerical results [see figure 4 of \cite{Hoye:2002at} for $a>5\,\mu$m, 
for example.]

Most experiments are carried out between a sphere (of radius $R$) and a plane.
In this circumstance, if $R\gg a$, $a$ being the separation between the
sphere and the plate at the closest point, the force may be obtained from
the proximity force approximation, 
\be
\mathcal{F}=2\pi R F(a), 
\ee
$F(a)$ being the free energy for the case of parallel plates separated
by a distance $a$.  Thus in the idealized description, the low temperature
dependence including our linear term is
\be
\fl\mathcal{F}\sim-\frac{\pi^3 R}{360 a^3}\left[1-\frac{45}{\pi^3}\zeta(3)aT
+\frac{360}{\pi^3}\zeta(3)(aT)^3-16(aT)^4\right],\qquad aT\ll1.
\ee
Since this conversion is trivial, in the following we will restrict attention
to the straightforward parallel plate situation.

These results are only approximate, because they assume the metal is ideal 
except for the exclusion of the TE zero.  Elsewhere, we have referred to
this model as the Modified Ideal Metal (MIM) model \cite{Hoye:2002at, 
Milton:2004ya}.  Evidently, for sufficiently low temperatures the
approximation used here, that $\zeta_1\gg\gamma$, breaks down, the 
function $f(m)$ becomes continuous, and the linear term disappears.
Indeed, numerical calculations based on real optical data for the
permittivity show this transition.  An example of such a calculation
is presented in \sref{sec:calc}.  There, in \fref{fig1}, we see a negative 
slope in the quantity $P/P_C$ as a function of the plate separation $a$ 
in the region between 1 and 2 micrometers.  This slope is approximately
$-0.1/\mu\mbox{m}$.  Here $P$ is the pressure between the plates at 300 K,
while $P_C$ is the ideal Casimir pressure \eref{caszero}.  If we compare
this to our approximate prediction \eref{linear},
\be
\frac{P^{T=300\,{\rm K}}}{P_C}\approx 1-\frac{30}{7.62}\frac{\zeta(3)}{\pi^3}
\frac{a}{\mu\mbox{m}}=1-0.15 \frac{a}{\mu\mbox{m}},
\ee
the slope and intercepts agree at the 20\% level.  Accurate numerical results
between real metal plates and sphere are given in \cite{Brevik:2004ue}.

Because this linear behavior does not persist at arbitrarily small 
temperatures,
it is clear that the conflict with the third law anticipated in the
arguments in the previous section do not apply.  In fact, as we shall now see,
the entropy does go to zero at zero temperature. 
   
\section{Arguments in favor and against the TE zero mode}
\label{Sec:arg}
As noted above, there are strong thermodynamic and electrodynamic arguments
in favor of the exclusion of the TE zero mode.  Essentially, the point
is that a realistic physical system can have only one state of lowest
energy. Electrodynamically, one can start from the Kramers-Kronig relation
that relates the real and imaginary part of the permittivity, required by
causality, which can be written in the form of a dispersion
relation for the electric susceptibility \cite{embook}
\be
\chi(\omega)=\frac{\omega_p^2}{4\pi}\int_0^\infty \rmd \omega'
\frac{p(\omega')}{\omega^{\prime 2}-(\omega+\rmi \epsilon)^2}.
\ee
If the spectral function $p(\omega')\ge0$ is nonsingular at the origin,
it is easily seen that $\omega^2\chi(\omega)\to0$ as $\omega\to 0$, which
as shown in the previous section implies the absence of the TE zero mode.
Conversely, $p(\omega')$ must have a $\delta$-function singularity at
the origin to negate this conclusion.  This would seem implausible for
any but an overly idealized model.  In contrast, in the Drude model
\be
p(\omega')=\frac2\pi\frac\gamma{\omega^{\prime2}+\gamma^2}\to2\delta(\omega')
\qquad\gamma\to0.
\ee

It has been objected that rather than employing bulk permittivities
as done in the usual expression for the Lifshitz formula, one should
use surface impedances instead \cite{geyer,bezerra04,Bezerra:2005hc,decca2}.
Indeed this may be done, but it leads to identical results.  The surface
impedance merely expresses the linear relation between tangential
components of the electric and magnetic fields at the interface between
the two media,
\be
\mathbf{E}_\perp=Z(\omega,\mathbf{k}_\perp) \mathbf{B_\perp\times n},
\qquad \mathbf{n\times E_\perp}=Z(\omega,\mathbf{k}_\perp)\mathbf{B}_\perp,
\ee
where $\bf n$ is the normal to the interface at the point in question.
From Maxwell's equations we deduce \cite{embook,Milton:2004ya} for the
reflection coefficient for the TE modes
\be
r^{\rm TE}=-\frac{\zeta+Z \kappa}{\zeta-Z\kappa},\qquad \kappa^2=
\zeta^2+k_\perp^2,
\ee
and the surface impedance is\footnote{Here we have assumed
that the permittivity is independent of transverse momentum.  In principle
this is incorrect, although optical data suggest that the transverse
momentum dependence of $\varepsilon$ is rather small. See also 
\cite{esquivel05}.}
\be
Z=-\frac{\zeta}{\sqrt{\zeta^2[\varepsilon(\rmi \zeta)-1]+\kappa^2}}.
\label{surfimp}
\ee
From this reflection coefficient the Lifshitz formula is constructed
according to $d=\left(r^{\rm TE}\right)^{-2}\rme^{2\kappa a}-1$.  Evidently
the resultant expression for the Lifshitz pressure coincides with
that found from the permittivity, seen for example in \eref{tesummand}.
This coincidence has been well recognized by previous authors \cite{bostrom00,
mochan}.  The reason why Mostepanenko and collaborators obtain a different
result is that they omit the transverse momentum dependence in \eref{surfimp}
and thereby argue that at zero frequency $Z$ vanishes, 
\be
Z\to -\frac1{\sqrt{\varepsilon(\rmi \zeta)}}\sim \frac{\sqrt{\gamma}}{\omega_p}
\sqrt{\zeta},
\ee
which is the content of the normal skin effect formula
\be
Z(\omega)=-(1-\rmi)\sqrt{\frac\omega{8\pi\sigma}}
\ee
where $\sigma$ is the conductivity. [These two formulas are seen to be
identical if we replace $\omega=\rmi \zeta$ and recognize that 
$\gamma=\omega_p^2/(4\pi\sigma)$.]  These formulas apply when we have
the restriction appropriate to real photons $k_\perp^2\le \omega^2$.  
However, no such mass-shell condition applies to the virtual or evanescent
photons involved in the thermal Casimir effect.  The same sort of error
seems to be made by Torgerson and Lamoreaux \cite{torgerson1,torgerson2},
and by Bimonte \cite{bimonte1,bimonte2}. 

As noted above, use of the plasma model in the reflection coefficients
would lead to the conventional temperature dependence, but this dispersion
relation is inconsistent with real data.  However, it has been argued that 
in the ideal Bloch-Gr\"uneisen model \cite{bloch} the relaxation parameter
goes to zero at zero temperature.  However, real metals exhibit scattering
by impurities; in any case, at sufficiently low temperatures the residual
value of the relaxation parameter does not play a role, as the frequency
characteristic of the anomalous skin effect becomes dominant \cite{svetovoy}.
Moreover, the authors of \cite{decca2, Bezerra:2005hc} also extrapolate the 
plasma formula from the infrared region down to zero frequency, whereas in fact
frequencies very small compared to the frequency corresponding to the
separation distance play a dominant role in the temperature dependence
\cite{svetovoy}.   Finally, we emphasize that all present experiments
are carried out at room temperature, where the known room temperature data
are relevant.

The principal reason for the theoretical controversy has to do with the
purported violation of the third law of thermodynamics if the TE zero mode
is not included.  If ideal metal reflection coefficients are used otherwise
(the MIM model) such a violation indeed occurs, because the free energy
per unit area for small temperature then behaves like
\be
F=F_0+T\frac{\zeta(3)}{16\pi a^2}.  
\ee
However, we and others have shown \cite{Brevik:2003rg,Milton:2004ya,
svetovoy,sernelius} that for real metals,
the free energy per area has a vanishing slope at the origin. Indeed, in the 
Drude model we have
\be
F=F_0+T^2 \frac{\omega_p^2}{48\gamma}(2\ln2-1),
\ee
for sufficiently low temperatures. There is, however, an intermediate range
of temperatures where it is expected that the entropy is negative.
We do not believe that this presents a thermodynamic difficulty, and reflects
the fact that the electrodynamic fluctuations being considered represent
only part of the complete physical system \cite{Brevik:2003rg,Hoye:2002at,
Brevik:2004ue}, although this is not a universal opinion \cite{svetovoy}.
(See also the further remarks in \sref{Sec:Concl}.)
New calculations are underway, showing explicitly the zero slope of the
curve for the free energy near $T=0$, thus corresponding to zero entropy
\cite{bha}.

Two other recent papers also lend support to our point of view.  Jancovici
and \v Samaj \cite{jancovici} and Buenzli and Martin \cite{buenzli}
have examined the Casimir force between ideal-conductor walls with
emphasis on the high-temperature limit.  Not surprisingly, ideal inert
boundary conditions are shown to be inadequate, and fluctuations within
the walls, modeled by the classical Debye-H\"uckel theory, determine the
high temperature behavior.  The linear in temperature  behavior
of the Casimir force is found to be reduced  by a factor of two
from the behavior predicted by an ideal metal, just as in \eref{hightlinear}.
  This is precisely the
signal of the omission of the $m=0$ TE mode.  Thus, it is very hard to
see how the corresponding modification of the low-temperature behavior
can be avoided.

Further support for our conclusions can be found in the  recent paper
of Sernelius \cite{sernelius05}, who calculates the van der Waals-Casimir
force between gold plates using the Lindhard or random phase approximation
dielectric function.  
The central theme of his work is to describe the thermal Casimir
effect in terms of {\it spatial dispersion}. Physically, spatially
nonlocal effects play a role at low frequencies because charge carriers
can move freely over large distances. Deviations from standard local
electromagnetic theory can then be expected. Spatial dispersion implies
that the standard Fresnel equations no longer apply. Moreover, because of
lack of experimental data one has to rely on theoretical model-dielectric
functions.

Sernelius finds \cite{sernelius05} that for large separations the force is 
one-half that of the ideal metal, just as in the calculation in 
Refs.~\cite{jancovici,buenzli}. This
agreement is not quite trivial; it means that the thermal Casimir effect
can be explained in two, apparently unrelated ways: One way is to
include spatial dispersion in the formalism from the beginning, omitting
dissipation. The other way is the conventional one, namely to describe the
thermal effect in terms of dissipation alone, by introducing the
relaxation frequency $\gamma$ as we have done above. Sernelius shows
that, for arbitrary separation between the plates, the spatial-dispersion
results nearly exactly coincide with the local dissipation-based results
\cite{sernelius01,bostrom}.

\section{Experimental constraints}
\label{Sec:exp}
We have marshaled theoretical arguments that seem to us quite overwhelming
in favor of the absence of the TE zero mode in the temperature dependence
of the Casimir force between real metal plates, which seem to imply
unambiguously  that there should be large ($\sim 15\%$) thermal corrections to 
the Casimir force at separations of order 1 micrometer.  New detailed 
calculations based on this theory, and using real optical data for aluminum, 
are discussed in the following section.  The difficulty is that, 
experimentally, it is not easy to perform Casimir force measurements at other 
than room temperature, so current constraints on the theory all come from room
temperature experiments.  Then all one can do is compare the theory at
room temperature with the experimental results, which must be corrected
for a variety of effects, such as surface roughness, finite conductivity,
and patch potentials.  A deviation between the corrected zero
temperature theory and the room temperature
observations then is taken as a measure of the temperature correction.

The temperature correction is evidently relatively largest at the largest
separations, where, unfortunately, the total Casimir force is weakest.  
Lamoreaux's
early experiments \cite{lamoreaux} were conducted at the 1 $\mu$m scale,
so if they were accurate to 10\% they would have seen the effect our
theory predicts, but probably, in spite of Lamoreaux's assertion, they were
not so accurate, because few essential corrections were included
\cite{mohideen99}.  The
experiments of Mohideen \etal \cite{mohideen, mohideen1, mohideen2} were
much more accurate, but because they were conducted at much smaller
distances, even our rather large temperature correction would have remained
inaccessible.  It is the most recent experiments of the Purdue group
\cite{decca1,decca2} that claim the extraordinarily high precision to
be able to see our effect at distances as small as 100 nm.  Indeed, they
see no deviation from the corrected
zero-temperature Lifshitz theory using optical
permittivities, and hence assert that our theory is decisively ruled out.
The effect we predict for the temperature correction is only 1.5\% at
a distance of 160 nm \cite{Hoye:2005eu}, so the measurement must be
performed at the 1\% level to see the effect there. (For the usually
employed sphere-plate configuration, $\Delta\mathcal{F}/\mathcal{F}\approx
2.5\%$ at $a=160$ nm.) Although they claim
this degree of accuracy, it is doubtful that they have achieved it,
because, for example, to achieve 1\% accuracy, the separation would
have to be determined to better than 0.3\%, or 0.5 nm at $a=160$ nm.
Since the roughness in the surfaces involved is much larger than this
(see also \cite{neto05}),
and other corrections (such as the fact that the metallic surfaces are
actually thin films, and the effects of surface
plasmons \cite{il,bordag05}) have not been included, we have reason to be
skeptical of such claims \cite{iannuzzi2}.

In any case, it would seem imperative to perform experiments at different
temperatures in order to provide evidence for or against temperature
dependence of Casimir forces. We understand such experiments are in
progress.  We encourage experimentalists to redouble their efforts to
determine the presence or absence of such an effect in an unbiased manner,
for the issues involved touch at the heart of our fundamental theoretical
understanding of electrodynamics, statistical mechanics, and quantum
field theory.

\section{New calculations}
\label{sec:calc}
\begin{figure}
\centering
  \includegraphics[width = 5in]{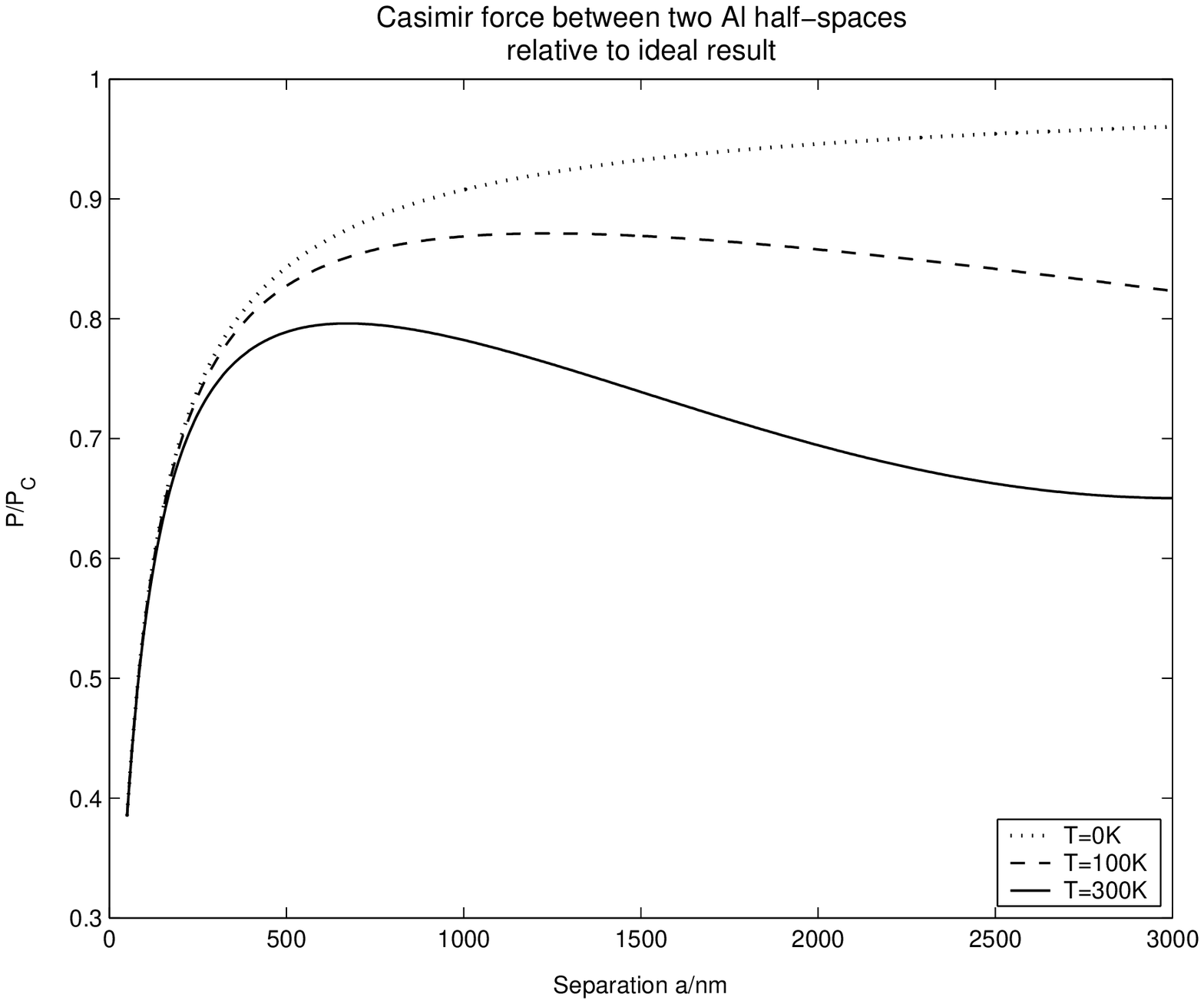}
\caption{\label{fig1} Temperature dependence of the Casimir force between 
aluminum plates.}
\end{figure}

\begin{figure}
\begin{center}
    \includegraphics[width = 5in]{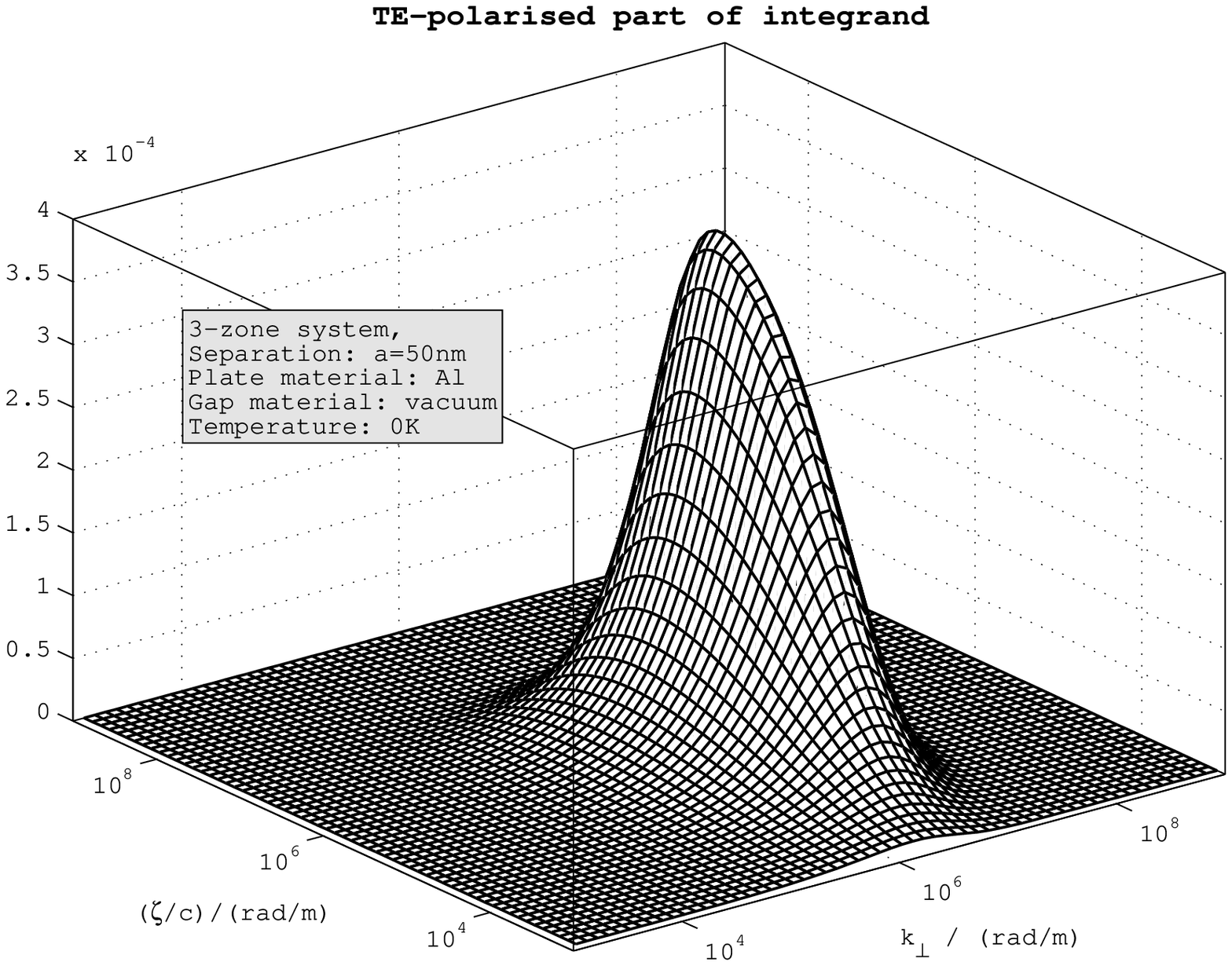}
\caption{\label{figint1}TE part of the integrand in \eref{integrands}.}
  \end{center}
\end{figure}

\begin{figure}
\begin{center}
    \includegraphics[width = 5in]{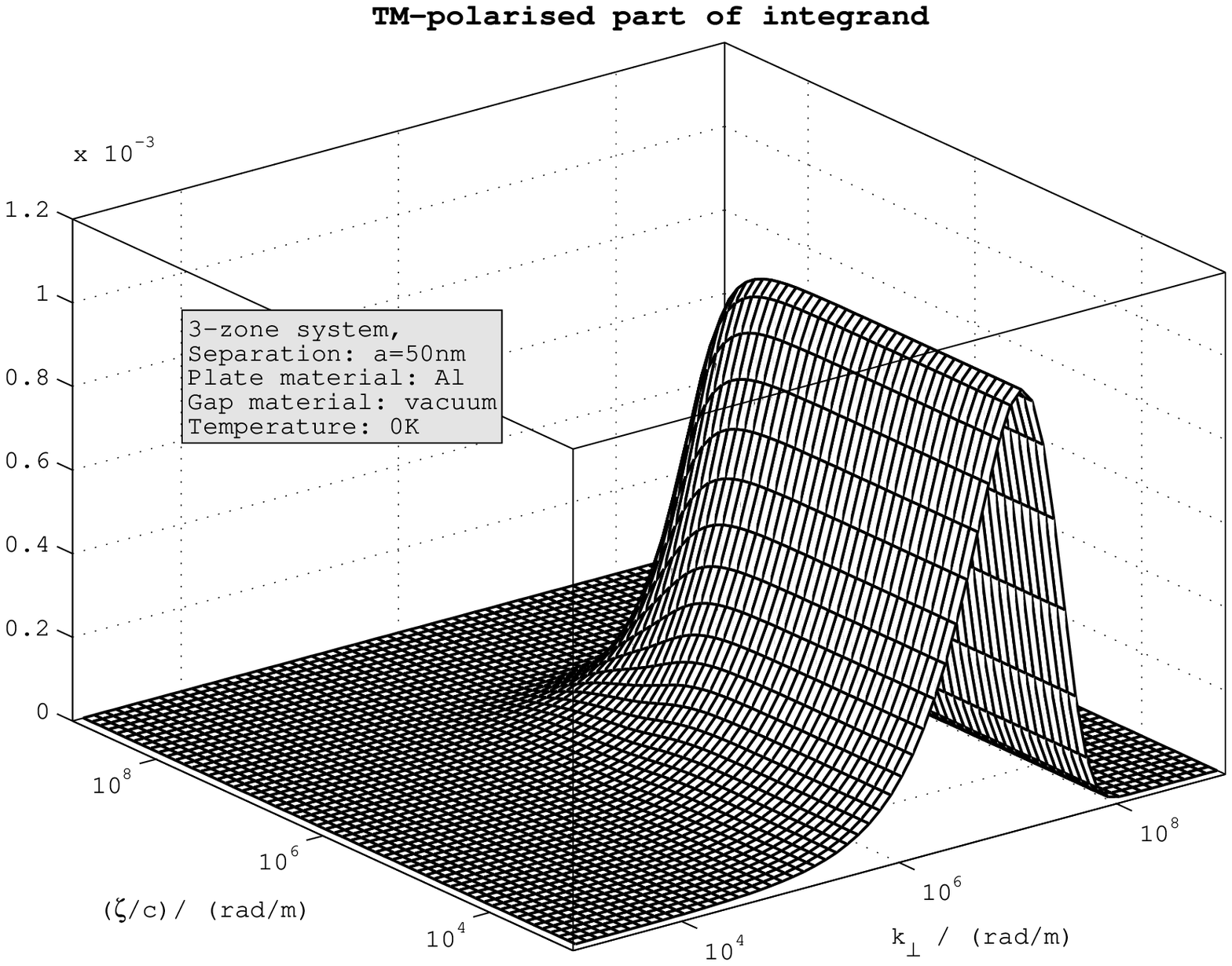}
\caption{\label{figint2}TM part of the integrand in \eref{integrands}.}
  \end{center}
\end{figure}

\begin{figure}
\begin{center}
    \includegraphics[width = 5in]{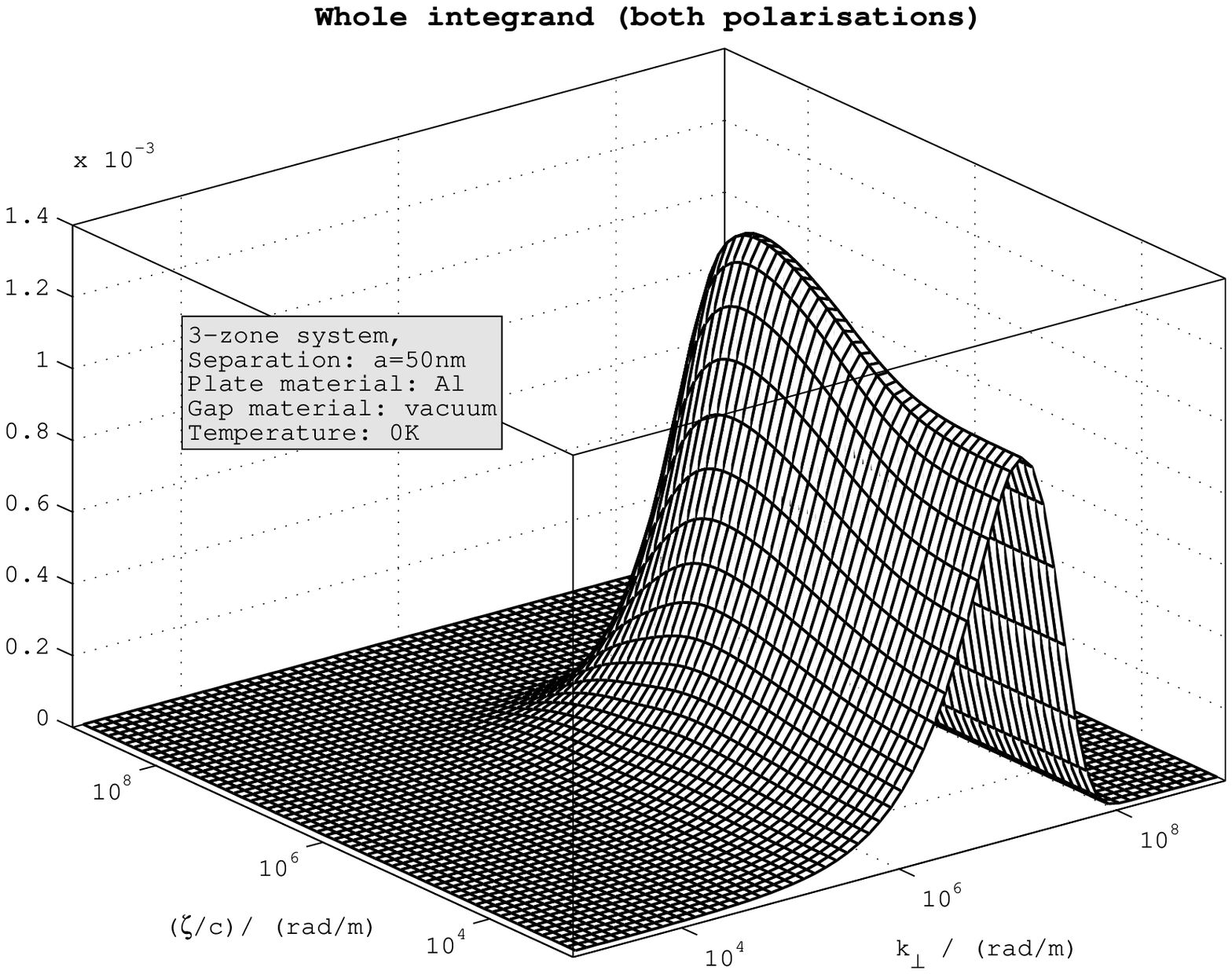}
\caption{\label{figint3}Total integrand in \eref{integrands}.}
  \end{center}
\end{figure}

\begin{figure}
\begin{center}
    \includegraphics[width = 5in]{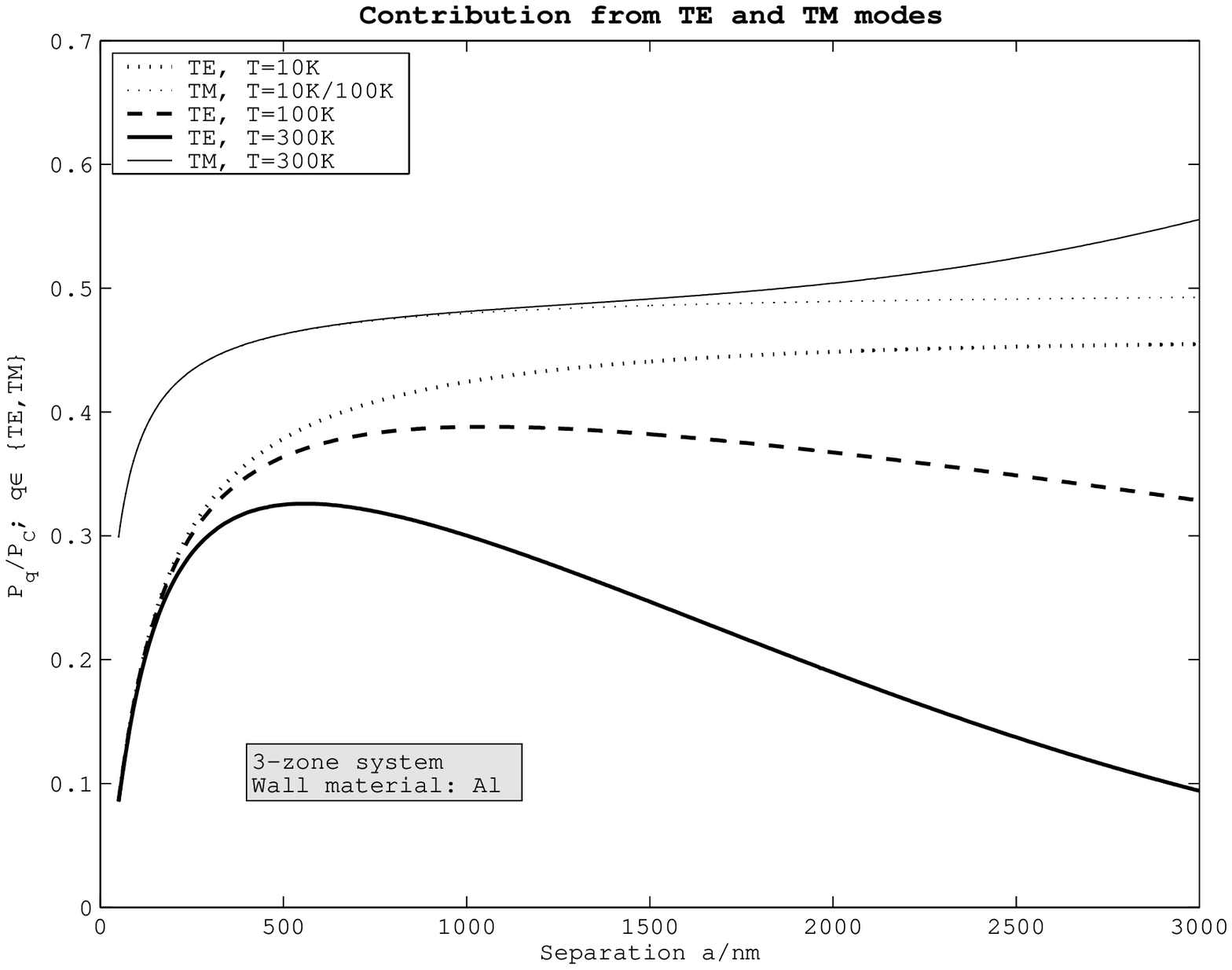}
\caption{\label{figa}TE and TM contributions to the pressure shown
in \fref{fig1}.}
  \end{center}
\end{figure}

\begin{figure}
\begin{center}
    \includegraphics[width = 5in]{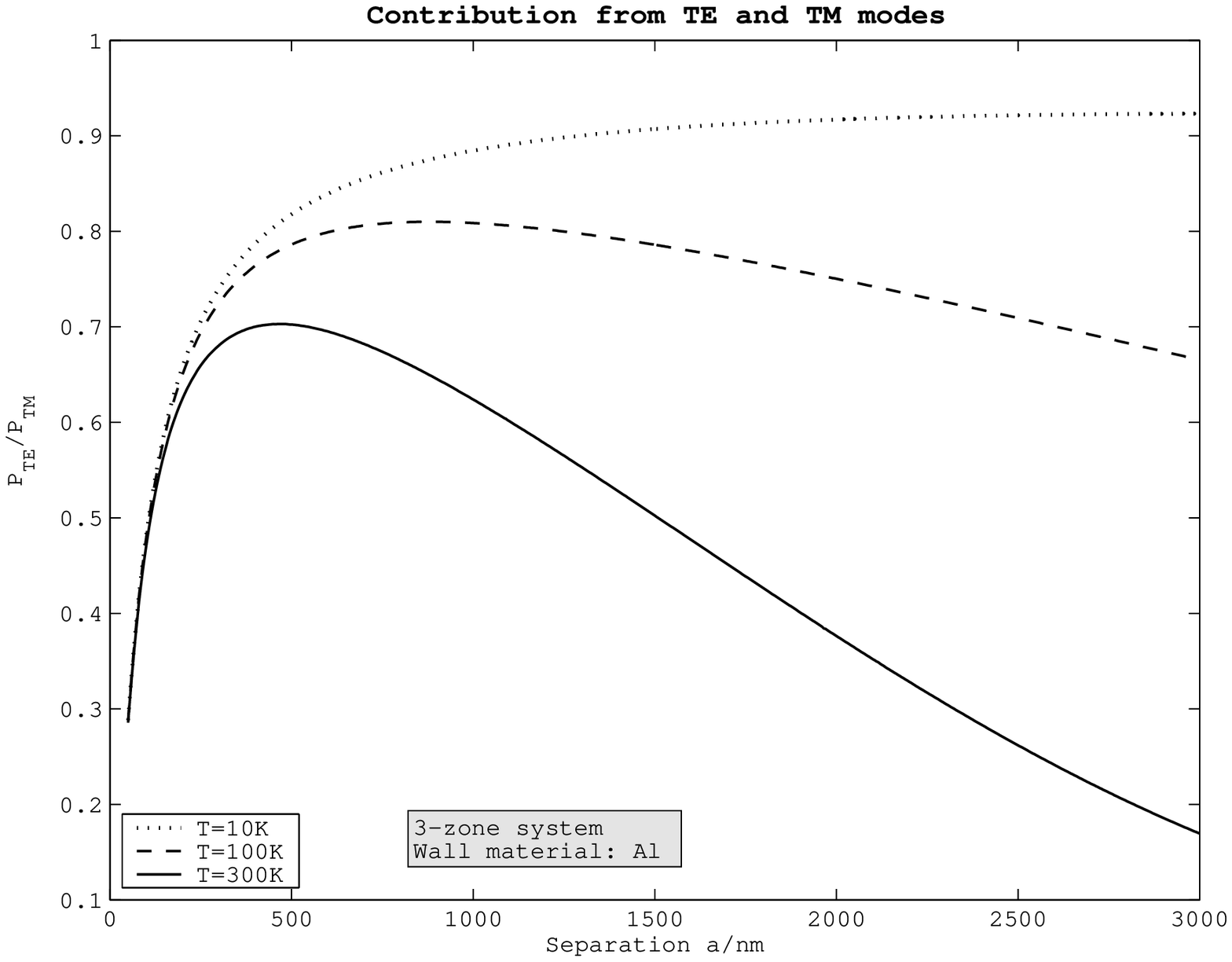}
\caption{\label{figb}Ratio of the TE and TM mode pressures contributing
to the pressure shown in \fref{fig1}.}
  \end{center}
\end{figure}

To aid in the experimental disentanglement of this effect,
we have carried out new calculations of the Casimir force between
two infinite half-spaces made of aluminum, separated by a vacuum space
of width $a$. (Other recent calculations appear in 
\cite{Hoye:2005eu, Bentsen:2005yi, ba}.)
 The results are shown in \fref{fig1}. (Figures \ref{fig1}--\ref{fig4}
are taken from the Master's Thesis of SAE \cite{ellingsen}.)
Formulas made use of are read off from \eref{lifshitz0t} and are now
given in usual dimensional units:
\numparts
\begin{eqnarray}
\fl P_{T=0}(a) = -\frac{\hbar}{2\pi^2}\int_0^\infty \rmd\zeta \int_0^\infty 
\rmd k_\perp k_\perp\kappa_0 \left(\frac{\Delta_{\rm TE}^2e^{-2\kappa_0a}}
{1-\Delta_{\rm TE}^2\rme^{-2\kappa_0a}}+\frac{\Delta_{\rm TM}^2e^{-2\kappa_0a}}
{1-\Delta_{\rm TM}^2e^{-2\kappa_0a}}\right),\\
\fl P_{T>0}(a) = -\frac{k_BT}{\pi}{\sum_{m=0}^{\infty}}{}^{\prime} 
\int_0^\infty \rmd k_\perp k_\perp\kappa_0 \left(\frac{\Delta_{\rm TE}^2
e^{-2\kappa_0a}}{1-\Delta_{\rm TE}^2e^{-2\kappa_0a}}+\frac{\Delta_{\rm TM}^2
e^{-2\kappa_0a}}{1-\Delta_{\rm TM}^2e^{-2\kappa_0a}}\right),
\end{eqnarray}
\endnumparts
where $a$ is the separation between plates (of infinite thickness), and
\numparts
\begin{eqnarray}
\Delta_{\rm TE} = \frac{\kappa - \kappa_0}{\kappa + \kappa_0},\label{deltate}
\\
\Delta_{\rm TM} = \frac{\kappa - \varepsilon(\rmi\zeta)\kappa_0}{\kappa + 
\varepsilon
(\rmi\zeta)\kappa_0},\label{deltatm}\\
\kappa = \sqrt{k_\perp^2 + \varepsilon(\rmi\zeta)\frac{\zeta^2}{c^2}}, \\
  \kappa_0 = \sqrt{k_\perp^2 + \frac{\zeta^2}{c^2}}.
\end{eqnarray}
\endnumparts
Here, $\varepsilon(\rmi \zeta)=\epsilon(\rmi \zeta)/\epsilon_0$ is the usual
permittivity relative to the vacuum.
In the case of finite temperatures, $\zeta=\zeta_m=2\pi m k_B T/\hbar$, and
a standard Lifshitz substitution of 
integration variables was made during calculations 
(see for example (3.2b) of \cite{Hoye:2002at}). 
The results are plotted relative to the standard 
Casimir pressure $P_C$ in \eref{caszero}.
Calculations have been carried to a relative accuracy of better than 
10$^{-4}$.  Even at $T=0$ there are large deviations from the ideal
Casimir result at all distance scales.

To illustrate the contributions of the TE and TM modes, 
figures \ref{figint1}--\ref{figint3} depict the TE and TM integrands
of a Casimir pressure expression of the type
\be
P_{T=0}=\int_0^\infty \rmd \zeta\int_0^\infty \rmd k_\perp [I_{\rm TE}
(\rmi \zeta,k_\perp)+I_{\rm TM}(\rmi \zeta,k_\perp)].\label{integrands}
\ee
It is clear that the TE term in the integrand falls off rapidly to zero as
$\zeta\to 0$ whereas the TM term remains finite.
The relative contributions to the pressure by the TE and TM modes
are illustrated in figures \ref{figa}--\ref{figb}.  Evidently, the contribution
of the TE mode rapidly decreases with increasing temperature and increasing
plate separation.

\subsection{Results for a five-layer model}
\begin{figure}
  \begin{center}
    \includegraphics[width=3in]{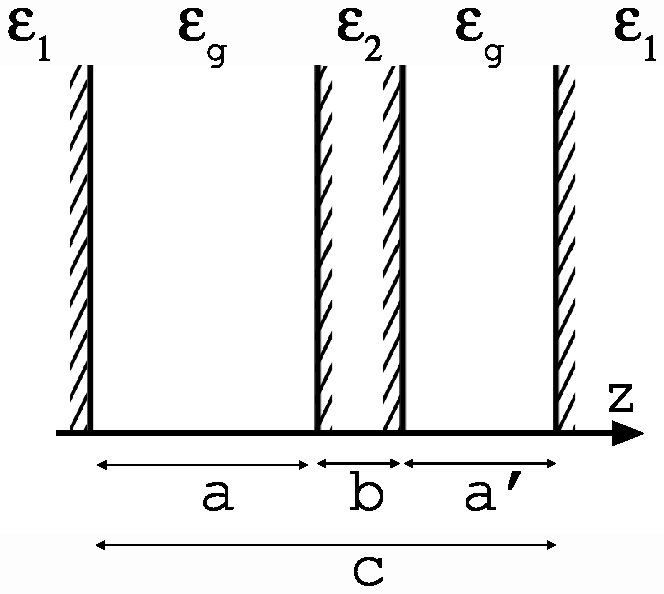}
\caption{\label{fig2} The 5-zone geometry. Here we have set 
$\varepsilon_g=1$.}
  \end{center}
\end{figure}

Because many experiments have been carried out with a conducting surface
between parallel capacitor plates it is useful to
consider the five layer geometry which has been treated repeatedly
by Toma\v s in recent years \cite{Tomas}.  The geometry is defined by
\fref{fig2}.  
All three slabs are assumed to be made of aluminum.
We assume one intermediate plate of width $b$, immersed in a cavity of
total width $c=a+a'+b$.  Between the central plate and the two outer
semi-infinite media we assume a vacuum ($\varepsilon_g=1$).  The quantity
$h$ is defined as $h=c-b$.  The quantity $\delta$ is the deviation of the
center of the plate from the midline of the cavity.  The Casimir pressure
at zero and finite temperature is given by
\numparts
\begin{eqnarray}
P_{T=0}(\delta; b,c)= \frac{\hbar}{2\pi^2}\int_0^\infty \rmd\zeta
\int_0^\infty \rmd k_\perp\sum_{q={\rm TE}}^{\rm TM}I_q(\rmi\zeta, k_\perp;
\delta,b,c)\\
P_{T>0}(\delta; b,c)=\frac{k_BT}{\pi}
{\sum_{m=0}^\infty}{}'\int_0^\infty \rmd k_\perp
\sum_{q={\rm TE}}^{\rm TM}I_q(\rmi\zeta_m, k_\perp;
\delta,b,c).
\end{eqnarray}
\endnumparts
Here the integrand is 
\begin{eqnarray}
I_q(\rmi\zeta, k_\perp;\delta,b,c)=k_\perp\kappa_0 2\Delta_{1q}
\Delta_{2q}(1-\rme^{-2\kappa_2b})\rme^{-\kappa_0h}\sinh 2\kappa_0\delta
\nonumber\\
\fl \times\left[-\Delta_{2q}^2\rme^{-2\kappa_2b} +1-
\Delta_{1q}^2\rme^{-2\kappa_0h}(\rme^{-2\kappa_2b}-\Delta_{2q}^2)
-2\Delta_{1q}\Delta_{2q}(1-\rme^{-2\kappa_2b})
\rme^{-\kappa_0h}\cosh 2\kappa_0\delta\right]^{-1},\nonumber\\
\end{eqnarray}
The Casimir force is positive for positive $\delta$, and is antisymmetric
around the cavity center $\delta=0$.
The index $q$ in the $\Delta$'s in the formulas runs over the polarizations
TE and TM, which are given by equations (\ref{deltate}), (\ref{deltatm}).
These results are equivalent to those found earlier by Toma\v s \cite{tomas02},
but presented in a more illustrative form.

\begin{figure}
  \begin{center}
    \includegraphics[width = 5in]{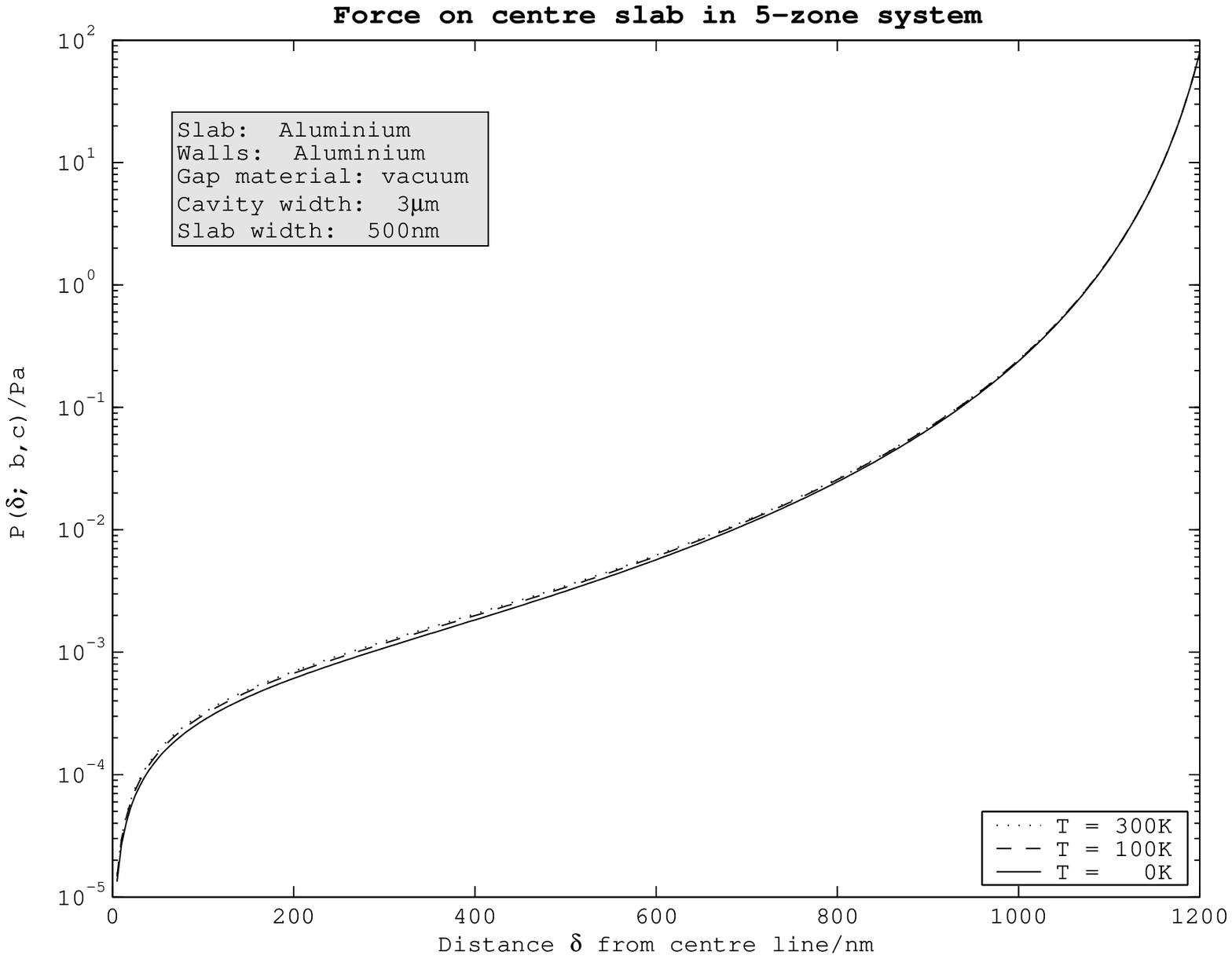}
    \caption{\label{fig3}
Pressure on the plate at a distance $\delta$ from the cavity
center. If $\delta<0$ one gets the antisymmetric prolongation about
the position $\delta = 0$.}
  \end{center}
\end{figure}

\begin{figure}
\begin{center}
    \includegraphics[width = 5in]{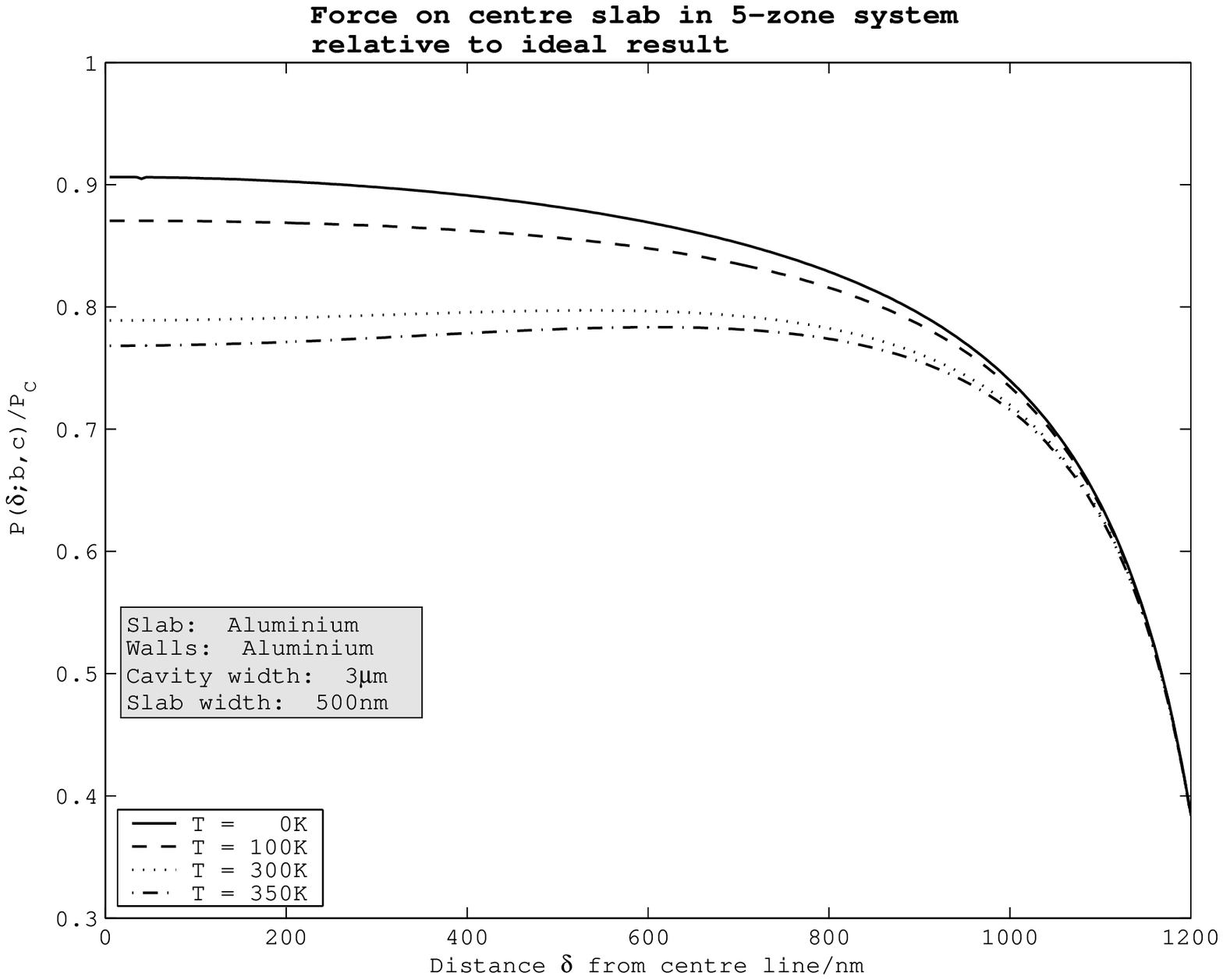}
\caption{\label{fig4}
Pressure on the plate at a distance $\delta$ from the cavity center
relative to Casimir's result \eref{cas} 
for ideal conductors. For $\delta<0$ one gets
the symmetric prolongation of the curve about the position $\delta = 0$.}
  \end{center}
\end{figure}

Numerically, we choose $c=3\,\mu$m and $b=500$ nm. The pressure
$P(\delta)$, calculated from $\delta=0$ to $\delta = 
(c-b)/2 - 50$ nm, is shown in \fref{fig3}.  
All calculations are done with an accuracy of better
that $10^{-4}$ in the final result, which should be sufficient for
practical purposes.
\Fref{fig4} shows the pressure relative to
Casimir's result for ideal conductors,
\begin{equation}
P_C=-\frac{\pi^2\hbar c}{240}\left(\frac{1}{(h/2 +\delta)^4}-
\frac{1}{(h/2 -\delta)^4} \right).\label{cas}
\end{equation}

As for the finite temperature calculations, it has been checked that all
terms in the sum (except for the zero mode) lie within the frequency domain
covered by Lambrecht's data.  There is thus no extra assumption made in 
the calculation, such as the property $\varepsilon\zeta^2/c^2\to0$
as $\zeta\to0$,  following from the Drude relation, except to 
ensure that there is no contribution from the
TE zero mode.  In the $T=0$ case, the Drude relation is used for low
frequencies.  The contribution to the force coming from
frequencies outside the Lambrecht region is very small,
smaller than the accuracy of the calculation.  Again, we
see large deviations from the ideal Casimir result at all temperatures,
as well as relatively large temperature corrections, which we hope will
be readily detectable in future experiments.

It would be very interesting to conduct experiments which are sensitive
to the 5-zone geometry we have considered, namely a slab in a cavity. 
In particular,
one could study the case of a slab oscillating about the center of a cavity, 
detecting how the oscillation frequency varies with temperature. Using a 
cavity of width 2--4 micrometers, say, one finds that the temperature
corrections are largest for slab-wall separations on the order of 1 micrometer.

\section{Conclusions}
\label{Sec:Concl}

We have shown how the Casimir pressure between parallel plates can be 
calculated with the inclusion, as well as with the exclusion, of the TE zero 
mode.  We emphasized (\sref{Sec:arg}) that there are strong thermodynamic and 
electrodynamic arguments in favour of the latter option. As is known in 
general, instead of employing bulk permittivities, it is possible to use 
surface impedances instead. We pointed out that in such a case it becomes 
necessary to take into account also the transverse momentum dependence in the 
expression \eref{surfimp} for the surface impedance. Otherwise, if one leaves
out the transverse momentum, one will obtain erroneous results as has often 
been the case in the literature 
\cite{decca2,Bezerra:2005hc,geyer,bezerra04,torgerson1,torgerson2,bimonte1,
bimonte2}.

We advocate the use of Drude's dispersion relation \eref{drude} throughout.  
The alternative plasma relation \eref{plasma} is inconsistent with real 
dispersive 
data. One peculiar effect arising from use of the Drude relation is the 
appearance of negative Casimir entropy in a finite frequency interval. The 
physical reason for this kind of behaviour is the following: It reflects the 
fact that we are dealing with only a part of the complete physical system. 
The effect may appear counterintuitive, but is not so uncommon in physics 
after all. One cannot apply usual thermodynamic restrictions such as 
positiveness of entropy to a ``subsystem'' formed by the induced interaction 
part of the free energy of the full system. This issue was discussed in detail 
in a previous paper (\cite{Hoye:2002at}).  In section IV of that paper we 
introduced, as an illustrative mechanical model, a system of two harmonic 
oscillators interacting via a third one. Such oscillators represent a 
simplified picture of two parallel plates interacting via the electromagnetic 
field. We calculated the classical as well as the quantum free energy of this 
mechanical system, and found there to be a finite temperature interval for 
which the interaction free energy increases with increasing temperature, thus 
leading to a negative interaction entropy term $S=-\partial F/ \partial T$. In 
this way a mechanical analogy with the Casimir interaction energy  (and 
corresponding interaction entropy) was demonstrated. 

The new developments of our paper are the establishment of the complete 
mathematical formalism necessary to calculate 
the temperature correction to the Casimir force. We assumed infinite, parallel,
plates, with no surface roughness included, and assumed in all calculations 
the tabulated dispersive data for bulk materials. Then,  we showed calculated 
results in figures \ref{fig1}--\ref{fig4} which are all new. They all assume 
aluminum plates. We gave results for three-layer geometry, and also some for 
a five-layer geometry. Our general suggestion for experimentally
measuring the influence from finite temperature is to keep 
the geometry of the setup as constant as possible (by using an invar 
material, for instance), and then measure the force at two accessible 
temperatures in the laboratory, for instance  300 K and 350 K.  It would
be even better if experiments could be carried out at liquid He temperatures;
we understand such experiments are underway.

\ack
KAM thanks the Physics Department of Washington University for its 
hospitality.  His work has been supported in
part by the US Department of Energy.  We are grateful to Bernard Jancovici,
Umar Mohideen and Roberto Onofrio for helpful conversations and correspondence,
and Astrid Lambrecht and Serge Reynaud for making their data available to us.

\end{document}